\documentclass[superscriptaddress,preprint]{revtex4}  
\usepackage{amsfonts}
\usepackage{amsmath}
\usepackage{cancel}
\usepackage{graphicx}
\usepackage{mathcomp}
\usepackage{color}
\usepackage{ulem}
\usepackage{xcolor} 
\usepackage{setspace}

\usepackage[utf8]{inputenc}
\usepackage{overpic}
\usepackage{chemarrow}

\newif\iffigures
\figurestrue

\renewcommand{\vec}{\mathbf}

\renewcommand{\refname}{REFERENCES}

\def\undertilde#1{\mathord{\vtop{\ialign{##\crcr
$\hfil\displaystyle{#1}\hfil$\crcr\noalign{\kern1.5pt\nointerlineskip}
$\hfil\widetilde{}\hfil$\crcr\noalign{\kern1.5pt}}}}}

\begin{document}
\title{Kinetic investigation of the planar Multipole Resonance Probe in the low-pressure plasma}

\author{Chunjie Wang}
\affiliation{Institute for Theoretical Electrical Engineering, Ruhr University Bochum, Center for Plasma Science and Technology, D-44780 Bochum, Germany}
\author{Michael Friedrichs}
\affiliation{South Westphalia University of Applied Science Soest, Department of Electrical Power Engineering, Lübecker Ring 2, 59494 Soest, Germany}
\author{Jens Oberrath}
\affiliation{South Westphalia University of Applied Science Soest, Department of Electrical Power Engineering, Lübecker Ring 2, 59494 Soest, Germany}
\author{Ralf Peter Brinkmann}
\affiliation{Institute for Theoretical Electrical Engineering, Ruhr University Bochum, Center for Plasma Science and Technology, D-44780 Bochum, Germany}

\date{\today}

\begin{abstract}
Active Plasma Resonance Spectroscopy (APRS) is a well-established plasma diagnostic method: a radio frequency signal is coupled into the plasma via a probe or antenna, excites it to oscillate, and the response is evaluated through a mathematical model. The majority of APRS probes are invasive and perturb the plasma by their physical presence. The planar Multipole Resonance Probe (pMRP) solves this problem: it can be integrated into the chamber wall and minimizes the perturbation. Previous work has studied the pMRP in the frame of the Drude model, but it misses important effects like collision-less damping. In this work, a collision-less kinetic model is developed to further investigate the behavior of the pMRP. This model consists of the Vlasov equation, which is coupled with the Poisson equation under electrostatic approximation. The spectral response of the probe-plasma system is found by calculating the complex admittance. This model covers the kinetic effects and overcomes the limitations of the Drude model.

\vspace{3mm}
\vspace{10mm}

\end{abstract}

\maketitle
\newpage

\section{Introduction}
\textit{Plasma resonance spectroscopy} (PRS) denotes a type of diagnostic method that utilizes the ability of plasma to resonate at or near the electron plasma frequency $\omega_\mathrm{pe}=\sqrt{e^2 n_\mathrm{e}/ \left(\epsilon_0 m_\mathrm{e}\right)}$. An overview and classification of PRS can be found in \cite{Lapke2011}. Here, we only focus on \textit{active plasma resonance spectroscopy} (APRS) \cite{Lapke2013,Oberrath2014-1}: RF signals are fed into the plasma via a probe or antenna, the plasma is excited to oscillate, and the spectral response is recorded. The plasma parameters, such as electron density and temperature, can be obtained from the measured spectra through a specific mathematical model \cite{Lapke2011,Lapke2013,Oberrath2014-1,Harp1964,Crawford1965,Arshadi2016,Arshadi2017,Haas2005,Xu2010,	Lapke2008,Oberrath2014-2,Oberrath2020,Oberrath2016,Oberrath2018,Friedrichs2018,Buckley1966}. 

Based on the concept of APRS, a variety of probes have been invented, such as the plasma resonance probe \cite{Harp1964,Crawford1965,Takayama1960}, curling probe \cite{Arshadi2016,Arshadi2017,Pandey2014}, hairpin probe \cite{Haas2005,Xu2010,Piejak2004,Piejak2005}, and Multipole Resonance Probe (MRP) \cite{Lapke2011,Lapke2008,Oberrath2014-2,Oberrath2020,Fiebrandt2017}. As an optimized realization of APRS, the MRP allows a unique resonance peak in its spectrum and a simple relation between the resonance and the electron plasma frequency. The MRP provides the possibility of simultaneously measuring electron density, electron temperature, and electron-neutral collision frequency, which makes it a promising candidate for the supervision and control of industrial plasma. However, most APRS probes, including the MRP, can lead to plasma perturbation due to their invasive probe tips and holders. Once the probe is immersed in the plasma, a sheath area with a thickness of several Debye lengths will appear around the probe tip and its holder. The plasma perturbation is inevitably caused. To minimize the influence of the probe on plasma, non-invasive probes are preferred. Therefore, based on the invasive design-MRP, the planar Multipole Resonance Probe (pMRP) was proposed \cite{Schulz2014-1}, which can be flatly integrated into the chamber wall.

To calculate plasma parameters from the measured spectra, the mathematical model is very important. An analytic model of electrostatic APRS was derived based on the Drude model (cold plasma model) \cite{Lapke2013}. In the Drude model, electrons gain energy from the electric field and lose energy via electron-neutral collisions. But the Drude model misses important kinetic effects like collision-less damping. To capture the kinetic effects, Oberrath and Brinkmann \cite{Oberrath2014-1} proposed a fully kinetic generalization of \cite{Lapke2013}, which is valid for arbitrary probe geometry and arbitrary pressure. In the kinetic model, the temporal variation of kinetic free energy is governed by the difference between the power input from electrodes, collisional loss (electron-neutral collisions), and collision-less loss (electron-affiliated energy outflow). In \cite{Oberrath2020}, Oberrath presented the first calculated kinetic spectra for a MRP without the dielectric. Strong kinetic damping was captured. The kinetic model is capable of covering both collisional damping and collision-less damping \cite{Oberrath2014-1,Oberrath2016,Oberrath2018,Oberrath2020}.  

Until now, all the presented pMRP spectra are based on the Drude model. In \cite{Schulz2014-2,Schulz2014-3,Pohle2018}, the spectral response was investigated through the 3D-electromagnetic field simulation with CST Microwave Studio. In \cite{Friedrichs2018}, the first analytic model was solved by the functional analytic approach. To capture the pure kinetic effects, we formulate a collision-less kinetic model in this paper. This model is valid in the low-pressure plasma (a few $\mathrm{Pa}$), where the collision-less kinetic damping is dominant and the collisional damping is negligible.

\pagebreak

\section{collision-less kinetic model of the probe-plasma system}
As shown in Fig.~\ref{Ideal pMRP}, the idealized pMRP consists of two semi-disc electrodes $E_\pm$ which are insulated from each other and from the grounded chamber wall. A thin dielectric with a thickness of $d$ covers the electrodes and the chamber wall. To allow for an analytic solution, the size of the chamber wall is assumed to be infinite and the insulator is negligible \cite{Friedrichs2018}. The interaction domain of the probe-plasma system is composed of the dielectric and the plasma (including the sheath). It is advantageous to employ a naturally oriented Cartesian coordinate system $(x,y,z)$. The boundary between the dielectric and the plasma is assumed at $z=0$. The dielectric domain is located at $-d<z<0$ and the plasma domain is located at the half-space $z>0$.

The RF voltages $V=\pm \hat{V} \cos(\omega t)$ are applied to the electrodes $E_\pm$ in an antisymmetric fashion. The voltage frequency $\omega$ is much higher than the ion plasma frequency $\omega_\mathrm{pi}$, but is comparable to the electron plasma frequency $\omega_\mathrm{pe}$. Therefore, the active region is much smaller than the wavelength, which means that the electrostatic approximation can be employed in the electric field calculation. It is advisable to use dimensionless notation: $\Phi \rightarrow (T_\mathrm{e}/e)\Phi$, $z \rightarrow \lambda_\mathrm{D}z$, $t \rightarrow t/\omega_\mathrm{pe}$, $v_\mathrm{e} \rightarrow \sqrt{T_\mathrm{e}/m_\mathrm{e}}v_\mathrm{e}$, $v_\mathrm{i} \rightarrow \sqrt{T_\mathrm{e}/m_\mathrm{i}}v_\mathrm{i}$, $n \rightarrow \hat{n} n$. Here, $\hat{n}$ denotes the electron density $n_\mathrm{e}|_{z \rightarrow \infty}$ in the Bohm model which will be introduced in the next section. Under the electrostatic approximation, the Poisson equation relates the electric potential to the charge density
\begin{equation}
-\nabla \cdot \left(\epsilon_\mathrm{r} \nabla \Phi \right)=
\begin{cases}
0                                  &\text{Dielectric} \\
n_\mathrm{i}-n_\mathrm{e}    \quad &\text{Plasma} 
\end{cases},
\label{potential-D&P}
\end{equation}
with $\epsilon_\mathrm{r}=\epsilon_\mathrm{D}$ in the dielectric and $\epsilon_\mathrm{r}=1$ in the plasma. 

In the low-pressure plasma (a few $\mathrm{Pa}$), all kinds of collisions are very weak. For example, the ion-neutral collision frequency is much smaller than the ion plasma frequency, which makes ion-neutral collisions negligible. Since the pMRP frequency is much higher than the ion plasma frequency, the ion's response to the RF electric field can be neglected. The ion follows the collision-free and chemistry-free cold ion model: the equation of continuity represents a constant ion flux
\begin{equation}
n_\mathrm{i}(z)\, v_\mathrm{i}(z)=-1, 
\label{ion continuity}
\end{equation}
and the equation of motion represents the ion acceleration through the electric field
\begin{equation}
v_\mathrm{i} \frac{\partial v_\mathrm{i}}{\partial z}=-\frac{\partial \Phi}{\partial z}. 
\label{ion motion} 
\end{equation}
In the low-pressure regime, the electron-neutral collision frequency is much smaller than the electron plasma frequency. Therefore, the electron dynamics are assumed to follow the Vlasov equation 
\begin{equation}
\frac{\partial f}{\partial t}+\vec{v} \cdot \nabla_r f+\nabla\Phi \cdot \nabla_v f=0.
\label{Vlasov}
\end{equation}
$f=f(\vec{r},\vec{v},t)$ is the electron distribution function which is based on the six-dimensional phase space $(x,y,z,v_x,v_y,v_z)$.

If the pMRP electrodes are grounded, a floating sheath will appear in front of the dielectric, which denotes the static equilibrium. When the electrodes are applied with RF voltages $V=\pm \hat{V} \cos(\omega t)$, a small perturbation will be superimposed on the static equilibrium. It is advantageous to adopt the linear response theory: an equilibrium plus a small perturbation, which will be discussed in the next two sections. Here, it is important to note that the ion's response to the RF electric field is neglected because $\omega \gg \omega_\mathrm{pi}$.

\pagebreak

\section{Equilibrium and unperturbed trajectory}
Assuming that the pMRP electrodes are grounded $\bar\Phi|_{z=-d}=0$, a static equilibrium will appear. The dielectric surface $z=0$ holds a ``floating potential'', which is negative with respect to the plasma bulk. A sheath region with a thickness of a few Debye lengths exists in front of the dielectric. In this paper, the electron's equilibrium is assumed to be the Maxwell distribution
\begin{equation}
\bar f(z,\vec{v})=f_\mathrm{M}(z,\vec{v})=\frac{1}{(2 \pi)^{3/2}} 
\exp\left(-\frac{1}{2}\left(v_x^2+v_y^2+v_z^2\right)+\bar{\Phi}(z)\right),
\end{equation}
and the corresponding electron density can be described by the Boltzmann relation
\begin{equation}
n_\mathrm{e}(z)=\exp\left(\bar\Phi(z)\right). \label{ne-Bohm}
\end{equation}
A collision-less planar sheath thus appears, which follows the well-known Bohm model \cite{Bohm1949}. The ion density is given by the equation of continuity \eqref{ion continuity} and the equation of motion \eqref{ion motion}, and the electron density is given by the Boltzmann relation \eqref{ne-Bohm}. The electric potential is related to the difference of ion density and electron density via the Poisson equation
\begin{equation}
-\frac{\mathrm{d}^2 \bar\Phi(z)}{\mathrm{d} z^2}=n_\mathrm{i}(z)-n_\mathrm{e}(z),
\end{equation}
which subjects to the floating boundary condition
\begin{equation}
\bar\Phi(0)=-\frac{1}{2} \ln \left( \frac{m_\mathrm{i}}{2 \pi m_\mathrm{e}} \right).
\end{equation}
The floating potential $\bar\Phi(0)=-4.68$ in argon. As $z$ increases from $0$ to $\infty$, the electric potential increases monotonically from $\bar\Phi(0)$ to $0$. 

The motion of the electron under the static potential $\bar\Phi$ is defined as the unperturbed trajectory. The uniformity of the potential in the $x$ and $y$ space contributes to the uniform motion in the $x$ and $y$ directions. In the $z$ direction, the electron follows
\begin{gather}
	\frac{\mathrm{d}z}{\mathrm{d}t}= v_z,\\
	\frac{\mathrm{d}v_z}{\mathrm{d}t}=\bar\Phi^\prime(z).
\end{gather}
Here, we define $\varepsilon_z$ as the electron's total energy in the $z$ direction
\begin{equation}
\varepsilon_z=\mathcal{E}_z (z,v_z) = \frac{1}{2} v_z^2 - \bar\Phi(z).
\end{equation}
If $\varepsilon_z>-\bar\Phi(0)$, the electrons can overcome the sheath potential and reach the dielectric with $v_z=-\sqrt{2(\varepsilon_z+\bar\Phi(0))}$. In fact, the number of these high-energy electrons is very small, which allows us to employ the ``reflecting wall" approximation: these energetic electrons are assumed to be elastically reflected on the surface of the dielectric.

Generally, when an electron moves along the unperturbed trajectory, $\varepsilon_z$ remains constant. As depicted in Fig.~\ref{Unperturbed trajectory}, an electron enters from infinity at an initial speed of $v_z=-\sqrt{2\varepsilon_z}$, and then experiences a turnaround or reflection:
\begin{itemize}
	\item if $\varepsilon_z<-\bar\Phi(0)$, the electron turns around at $z={\bar\Phi}^{-1}\left(-\varepsilon_z\right)$;
	\item if $\varepsilon_z=-\bar\Phi(0)$, the electron reaches the dielectric with $v_z=0$, and then is dragged back into the plasma by the electric field;
	\item if $\varepsilon_z>-\bar\Phi(0)$, the electron is elastically reflected at $z=0$, 
	\begin{align}
	v_z:\,-\sqrt{2(\varepsilon_z+\bar\Phi(0))} \rightarrow \sqrt{2(\varepsilon_z+\bar\Phi(0))}.
	\end{align}
\end{itemize}
Finally, the electron leaves at a speed of $v_z=\sqrt{2\varepsilon_z}$.

For each coordinate pair $(z, v_z)$, we can calculate the time that is required for the electron to reach the turning point $(v_z<0)$ or the time that has elapsed since the electron left the turning point $(v_z>0)$. Assigning a negative sign to the first case and a positive sign to the second case, $\tau$ follows
\begin{equation}
\tau=\mathcal{T}(z,v_z) = \mathrm{sign}(v_z) \int_{z_\mathrm{min}}^z \frac{1}{\sqrt{v_z^2 - 2\bar\Phi(z) + 2\bar\Phi(z^\prime)}} \,\mathrm{d} z^\prime,
\end{equation}
in which 
\begin{equation}
z_\mathrm{min}=
\begin{cases}
\displaystyle{{\bar\Phi}^{-1}\left( -\frac{1}{2} v_z^2 + \bar\Phi(z) \right)} \quad  & \displaystyle{\frac{1}{2}v_z^2-\bar\Phi(z) < -\bar\Phi(0)} \\[2mm]
0    &\displaystyle{\frac{1}{2}v_z^2-\bar\Phi(z) \geq -\bar\Phi(0)}
\end{cases}.
\end{equation}

Based on the unperturbed trajectory, we can define a new coordinate system $(\varepsilon_z,\tau)$. $\varepsilon_z$ is the electron's total energy in the $z$ direction which remains constant along the unperturbed trajectory. $\tau$ describes the temporal parametrization of the unperturbed trajectory by selecting a turnaround or reflection as its reference point. The coordinate transformation is defined as follows:
\begin{equation}
(z, v_z) \, \autorightleftharpoons{$\varepsilon_z=\mathcal{E}_z(z, v_z), \tau=\mathcal{T}(z, v_z)$}{$z=Z(\varepsilon_z,\tau), v_z=V_z(\varepsilon_z,\tau)$} \, (\varepsilon_z,\tau).
\end{equation}
The functions $\mathcal{E}_z(z, v_z)$ and $\mathcal{T}(z, v_z)$ describe a regular coordinate transformation from the Cartesian coordinate system $(z,v_z)$ to the newly defined coordinate system $(\varepsilon_z,\tau)$. The inverse transformation functions $Z(\varepsilon_z,\tau)$ and $V_z(\varepsilon_z,\tau)$ are given by the specific solution of the equation of motion: the electron moves with a constant energy $\varepsilon_z$ and undergoes a turnaround or reflection at $\tau=0$. This inverse transformation can only be constructed numerically.

\pagebreak

\section{Linearized kinetic model}
In the previous section, the equilibrium situation is analyzed. When the pMRP electrodes are grounded, a planar sheath appears in front of the dielectric. If we apply the RF signals $V=\pm \hat{V} \cos(\omega t)$ to the electrodes, a perturbation will be superimposed on the equilibrium. Since the RF voltages are very small, the linear response theory applies \cite{Krall1973,Buckley1966}: the electron distribution function and potential can be expressed in terms of an equilibrium value plus a small perturbation
\begin{gather}
f(\vec{r},\vec{v},t)=\bar f(z,\vec{v}) \left( 1+\delta\!f(\vec{r},\vec{v},t) \right),
\label{linearization-f}\\
\Phi(\vec{r},t)=\bar \Phi(z)+\delta \Phi(\vec{r},t),
\label{linearization-Phi}
\end{gather}
with $\left| \bar f\,\delta\!f \right| \ll \left| \bar f \right|$ and $\left| \delta\Phi \right| \ll \left| \bar\Phi \right|$. Substituting \eqref{linearization-f} and \eqref{linearization-Phi} into the Vlasov equation \eqref{Vlasov} and Poisson equation \eqref{potential-D&P}, we obtain the linearized perturbation equations
\begin{gather}
\frac{\partial \delta\!f}{\partial t}+\vec{v} \cdot \nabla_r \delta\!f -\vec{v} \cdot \nabla \delta\Phi + \bar\Phi^\prime(z) \frac{\partial \delta\!f}{\partial v_z}=0, 
\label{f1-perturbation}\\
-\nabla \cdot (\epsilon_\mathrm{r} \nabla \delta\Phi)=
\begin{cases}
0                                                 &\text{Dielectric} \\
\displaystyle{-\int \bar f\,\delta\!f\,{\mathrm{d}^3 v}}  \quad &\text{Plasma}  
\end{cases}.
\label{phi1-perturbation}
\end{gather}
We assume that all perturbation quantities are time-harmonic according to
\begin{gather}
\delta\!f(\vec{r},\vec{v},t)=\mathrm{Re}\left[\delta\!\tilde{f}(\vec{r},\vec{v}) \exp(i\omega t)\right],\\
\delta\Phi(\vec{r},t)=\mathrm{Re}\left[\delta\tilde{\Phi}(\vec{r}) \exp(i\omega t)\right].
\end{gather} 
Considering the uniformity of the static equilibrium in the $x$ and $y$ space, we perform the Fourier transform
\begin{gather}
\delta\!\underline{\tilde{f}}(k_x,k_y,z,\vec{v})=\int_{-\infty}^{\infty} \int_{-\infty}^{\infty} \delta\!\tilde{f} (\vec{r},\vec{v}) \exp\left(i \left( k_x x+k_y y \right)\right) \mathrm{d}x \,\mathrm{d}y,\\
\delta \underline{\tilde{\Phi}}(k_x,k_y,z)=\int_{-\infty}^{\infty} \int_{-\infty}^{\infty} \delta\tilde{\Phi}(\vec{r}) \exp\left(i \left( k_x x+k_y y \right)\right) \mathrm{d}x \,\mathrm{d}y.
\end{gather}
The partial derivatives are thus simplified: $\displaystyle{\frac{\partial\,}{\partial x} \rightarrow -i k_x}$ ,$\displaystyle{\frac{\partial\,}{\partial y} \rightarrow -i k_y}$. Before performing the inverse Fourier transform, both $k_x$ and $k_y$ can be regarded as constants. Therefore, in the following derivation, $k_x$ and $k_y$ are temporarily omitted from $\delta\!\underline{\tilde{f}}$ and $\delta \underline{\tilde{\Phi}}$. The perturbation equations \eqref{f1-perturbation} and \eqref{phi1-perturbation} become
\begin{gather}
i(\omega - k_x v_x - k_y v_y) \delta\!\underline{\tilde{f}}+v_z \frac{\partial \delta\!\underline{\tilde{f}}}{\partial z}+ \bar\Phi^\prime(z)\frac{\partial \delta\!\underline{\tilde{f}}}{\partial v_z}+i(k_x v_x+k_y v_y)\delta\underline{\tilde{\Phi}} -v_z \frac{\partial \delta\underline{\tilde{\Phi}}}{\partial z}=0,
\label{kinetic-fourier1}\\
\left(k_x^2+k_y^2\right)\delta\underline{\tilde{\Phi}}-\frac{\partial^2 \delta\underline{\tilde{\Phi}}}{\partial z^2}=
\begin{cases}
0             &\text{Dielectric} \\
\displaystyle{-\int \bar f \,\delta\!\underline{\tilde{f}}\,\mathrm{d}^3 v}  \quad &\text{Plasma} 
\end{cases}.
\label{Poisson-fourier1}
\end{gather}
Performing the coordinate transformation $(z,v_z) \to (\varepsilon_z,\tau)$, the kinetic equation \eqref{kinetic-fourier1} becomes
\begin{equation}
i\left(\omega - k_x v_x - k_y v_y\right) \delta\!\underline{\tilde{f}}+\frac{\partial \delta\!\underline{\tilde{f}}}{\partial \tau} + i\left(k_x v_x+k_y v_y\right) \delta\underline{\tilde{\Phi}}-\frac{\partial \delta\underline{\tilde{\Phi}}}{\partial \tau}=0.
\end{equation}
By integrating along the unperturbed trajectory, we obtain the solution of the kinetic equation
\begin{equation}
\begin{gathered}
\delta\!\underline{\tilde{f}}\left(Z\left(\varepsilon_z,\tau\right),v_x,v_y,V_z\left(\varepsilon_z,\tau\right)\right)=\delta\underline{\tilde{\Phi}}(Z(\varepsilon_z,\tau))\\
-i\omega \int_{-\infty}^{\tau} \exp\left( i\left(\omega - k_x v_x - k_y v_y\right)\left(\tau^{\prime}-\tau\right) \right) \delta\underline{\tilde{\Phi}}\left(Z(\varepsilon_z,\tau^{\prime})\right) {\mathrm{d} \tau^{\prime}}.
\end{gathered}
\label{solution of kinetic equation}
\end{equation} 
In the solution of the kinetic equation, the first term denotes the static response in the probe-plasma interaction, the second term denotes the dynamic response. The perturbation generated by the RF electric field is accumulated along the unperturbed trajectory via the integral from $-\infty$ to $\tau$.
Substituting \eqref{solution of kinetic equation} into \eqref{Poisson-fourier1}, the Poisson equation finally yields an integro-differential equation
\begin{equation}
k^2 \delta\underline{\tilde{\Phi}}(z) -\frac{\partial^2 \delta\underline{\tilde{\Phi}}(z) }{\partial z^2}=
\begin{cases}
0      &\text{Dielectric} \\
\displaystyle{-\exp\left(\bar \Phi (z) \right) \delta\underline{\tilde{\Phi}}(z) + i\omega \int_0^\infty K\left(\omega,k,z,z^\prime\right) \delta\underline{\tilde{\Phi}}\left(z^\prime\right) \mathrm{d}z^\prime} &\text{Plasma}  
\end{cases},
\label{Poisson-F}
\end{equation}
where $k=\sqrt{k_x^2+k_y^2}$ and $K\left(\omega,k,z,z^\prime\right)$ is derived from the integral term in \eqref{solution of kinetic equation}. Due to the uniformity of the static equilibrium in the $x$ and $y$ space, the perturbation equations \eqref{kinetic-fourier1} and \eqref{Poisson-fourier1} show the symmetry between $k_x$ and $k_y$. Therefore, $k=\sqrt{k_x^2+k_y^2}$ finally appears in the Poisson equation \eqref{Poisson-F}. The corresponding boundary conditions are
\begin{gather}
\delta\underline{\tilde{\Phi}}(-d_+) =\delta\underline{\tilde{\Phi}}_\mathrm{p\!M\!R\!P},\\
\delta\underline{\tilde{\Phi}}(0_-) =\delta\underline{\tilde{\Phi}}(0_+),\\
\epsilon_\mathrm{D} \delta\underline{\tilde{\Phi}}^{\prime}(0_-) =\delta\underline{\tilde{\Phi}}^{\prime}(0_+),\\
\delta\underline{\tilde{\Phi}}(+\infty) =0 ,
\end{gather}
in which $\delta\underline{\tilde{\Phi}}_\mathrm{p\!M\!R\!P}$ denotes the input signal of the pMRP
\begin{equation}
\delta\underline{\tilde{\Phi}}_\mathrm{p\!M\!R\!P}=\hat{V} \iint_{E_\pm}  \mathrm{sign} (y) \exp\left(i (k_x x+k_y y)\right)\mathrm{d}x \, \mathrm{d}y .
\end{equation}

\subsection{Solution of the Poisson equation}
In the plasma domain, the Poisson equation \eqref{Poisson-F} yields an integro-differential equation whose analytical solution is inaccessible. For numerical implementation, we define the electric potential $(z\geq -d)$ as
\begin{equation}
\delta\underline{\tilde{\Phi}}(z) = \sum_{n=N_1}^{N_2} \delta\underline{\tilde{\Phi}}_n \left(H((n-1)\Delta z)-H(n\Delta z)\right),
\end{equation}
where $H(z)$ is the unit step function and $\Delta z$ is the grid size of the coordinate $z$. $\delta\underline{\tilde{\Phi}}_{N_1}$ represents the input voltage of the probe, and $\delta\underline{\tilde{\Phi}}_{N_2}$ represents that the potential decays to $0$ when it gets far away from the probe. The Poisson equation and the corresponding boundary conditions can thus be discretized into a matrix equation
\begin{equation}
\mathbf{A} \, \mathbf{\delta\underline{\tilde{\Phi}}}=\mathbf{b},
\label{matrix-eq}
\end{equation}
in which
\begin{equation}
\mathbf{\delta\underline{\tilde{\Phi}}}=\left( \delta\underline{\tilde{\Phi}}_{N_1} \quad \delta\underline{\tilde{\Phi}}_{N_1+1} \ \cdots \ \delta\underline{\tilde{\Phi}}_{-1} \quad \delta\underline{\tilde{\Phi}}_{0} \quad \delta\underline{\tilde{\Phi}}_{1} \ \cdots \ \delta\underline{\tilde{\Phi}}_{N_2-1}  \quad \delta\underline{\tilde{\Phi}}_{N_2} 
\right)^{T}.
\end{equation}
By solving the matrix equation, the numerical solution of the potential can be obtained. Here, we need to note that the matrix equation and its solution are the functions of $k_x$ and $k_y$. Hence $\displaystyle{\frac{\partial\delta\underline{\tilde{\Phi}}(k_x,k_y,z)}{\partial z}\bigg|_{z=-d_+}}$ can be approximately expressed by
\begin{equation}
\frac{\partial\delta\underline{\tilde{\Phi}}(k_x,k_y,z)}{\partial z}\bigg|_{z=-d_+} \approx 
\frac{\delta\underline{\tilde{\Phi}}_{N_1+1}(k_x,k_y) - \delta\underline{\tilde{\Phi}}_{N_1}(k_x,k_y) }{\Delta z}.
\label{eq:z=-d}
\end{equation}

\subsection{General admittance of the probe-plasma system}
The ratio of the RF current $I$ to the positive electrode voltage is defined as the general complex admittance of the probe-plasma system. The current is counted as positive when it flows from the electrode to the plasma. Due to the electrical antisymmetry of the pMRP, the current, that flows from the negative electrode to the plasma, should be $-I$ accordingly. Since the electrodes are shielded by the dielectric layer, the electrodes' conductive current density is equal to the displacement current density in the dielectric. Therefore, the general complex admittance can be explicitly expressed as 
\begin{equation}
Y(\omega)=-\frac{i \omega \epsilon_\mathrm{D}}{V}\int\!\!\!\!\int_{E_+} \frac{\partial{\delta\tilde{\Phi}(x,y,z)}}{\partial z} \bigg|_{z=-d_+} \mathrm{d}x \,\mathrm{d}y,
\end{equation}
where $\displaystyle{\frac{\partial{\delta\tilde{\Phi}(x,y,z)}}{\partial z}\bigg|_{z=-d_+}}$ is given by the inverse Fourier transform of \eqref{eq:z=-d}.

\pagebreak

\section{Spectral response}
The spectral response of the probe-plasma system can be expressed by the real part of the general complex admittance, and its theoretical derivation has already been introduced in the previous sections. In this section, these calculations are carried out for a pMRP with electrode radius $R=5\,\mathrm{mm}$, dielectric thickness $d=0.04 \,\mathrm{mm}$, and dielectric permittivity $\epsilon_\mathrm{D}=4.82$. 

Fig.~\ref{fig-kinetic} presents the first calculated kinetic spectrum of the pMRP, in which the probe monitors an argon plasma with electron density $n_\mathrm{e}=4\times10^{15} \,\mathrm{m}^{-3}$ and electron temperature $T_\mathrm{e}=2\,\mathrm{eV}$. The spectral resonance and its broadening by the collision-less kinetic damping are clearly visible. The real part of the admittance reaches its maximum value of $1.64\, \mathrm{mS}$ at $\omega=0.50\,\omega_\mathrm{pe}$; the half-width of the resonance peak is $0.19\,\omega_\mathrm{pe}$. To further analyze the kinetic spectral response, we compare it with the Drude model.

In the Drude model \cite{Lapke2013}, the potential follows the Poisson equation
\begin{equation}
-\nabla \cdot \left(\epsilon_\mathrm{0} \epsilon_\mathrm{r} \nabla \Phi \right)=0,
\label{Poisson-Drude model}
\end{equation} 
with the relative permittivity as
\begin{equation}
\epsilon_\mathrm{r}=
\begin{cases}
\epsilon_\mathrm{D}    &\text{Dielectric}\\
1     &\text{Sheath}\\
\displaystyle{1-\frac{\omega_\mathrm{pe}^2}{\omega(\omega-i \nu)}}    \quad &\text{Plasma bulk}
\end{cases}.
\end{equation}
By performing the Fourier transform on the $x$ and $y$ space, the Poisson equation \eqref{Poisson-Drude model} can be simplified into a second-order ordinary differential equation of variable $z$, which is easy to solve. 

In Fig.~\ref{fig-Drude}, the Drude model is solved when electron density $n_\mathrm{e}=4\times10^{15} \mathrm{m}^{-3}$, sheath thickness $\delta=0.5\,\mathrm{mm}$, and electron-neutral collision frequency $\nu=0.01\sim0.2\,\omega_\mathrm{pe}$. As $\nu$ varies, the spectra show a constant resonance frequency, which is identical to the kinetic model. This proves that both the Drude model and the kinetic model can provide an accurate resonance frequency for calculating the electron density from the measurement. In the kinetic model, the perturbation generated by the pMRP is accumulated along the unperturbed trajectory. When electrons leave the probe-plasma interaction domain, this accumulated energy will be taken away. Hence, collision-less damping appears. But the Drude model only covers collisional damping. As the collision frequency decreases, collisional damping decreases, which yields a smaller half-width and a higher amplitude in the spectrum. The spectra in Fig.~\ref{fig-Drude} get the same half-width as Fig.~\ref{fig-kinetic} at $\nu=0.081\,\omega_\mathrm{pe}=2.89\times10^8\,\mathrm{s}^{-1}$ and the same amplitude at $\nu=0.137\,\omega_\mathrm{pe}=4.89\times10^8\,\mathrm{s}^{-1}$. These two collision frequencies indicate that collision-less kinetic damping is non-negligible, and even plays a dominant role in the low-pressure plasma.

Via the Drude model, we can only calculate the electron density. Whereas, the kinetic model makes it possible to acquire both electron density and electron temperature from the measurement. In Fig.~\ref{fig-kinetic-Te}, the kinetic spectra are calculated from $1\,\mathrm{eV}$ to $5\,\mathrm{eV}$. As the electron temperature increases, the spectral resonance frequency increases, the half-width increases, and the amplitude decreases. In the Drude model, with $3\lambda_\mathrm{D}$ sheath thickness, the spectrum shows an almost identical resonance frequency to the kinetic model. However, it is impossible to locate a collision frequency where the half-width and amplitude arrive at an appropriate match simultaneously. Thus we define two collision frequencies: the spectrum of the Drude model holds the same half-width as the kinetic model at $\nu_\mathrm{\!_W}$, and the same resonance amplitude at $\nu_\mathrm{\!_A}$. For example, when $n_\mathrm{e}=4\times10^{15} \mathrm{m}^{-3}$ and $T_\mathrm{e}=2\,\mathrm{eV}$, $\nu_\mathrm{\!_W}=0.081\,\omega_\mathrm{pe}$ and $\nu_\mathrm{\!_A}=0.137\,\omega_\mathrm{pe}$. These two collision frequencies can provide an approximate range for the collision-less kinetic damping. As shown in Fig.~\ref{fig-nu}, both $\nu_\mathrm{\!_W}$ and $\nu_\mathrm{\!_A}$ increase as the electron temperature rises, which indicates that the collision-less kinetic damping increases as the electron temperature rises. Electrons enter the probe-plasma interaction domain, pick up energy from the probe, and then take the energy away. The electron-affiliated energy outflow finally leads to the collision-less kinetic damping \cite{Oberrath2014-1}. At higher temperatures, electrons can move faster, so more energy flows out in unit time. As the electron temperature increases, the collision-less kinetic damping gets stronger.

\pagebreak

\section{Conclusion and outlook}
In this paper, a collision-less kinetic model is developed to investigate the behavior of the idealized planar Multipole Resonance Probe in the low-pressure plasma. This is the first kinetic model of the pMRP's specific geometry. Under the collision-less assumption, the electron dynamics follow the Vlasov equation, which is coupled to the Poisson equation under the electrostatic approximation. If the pMRP electrodes are grounded, a floating sheath will appear in front of the dielectric, which follows the Bohm sheath model. When the RF voltages are applied to the electrodes, a small perturbation will be superimposed on the static equilibrium. The linear response theory is thus applicable: an equilibrium plus a small perturbation, which are discussed in this paper.

By calculating the real part of the general complex admittance, the spectral response of the probe-plasma system is obtained. The expected pure ``kinetic effects'' are captured. The spectral resonance and its broadening by the collision-less kinetic damping are clearly visible. The collision-less kinetic damping is non-negligible, and even plays a dominant role in the low-pressure plasma. As the electron temperature increases, the kinetic damping gets stronger. In the low-pressure plasma, by covering the collision-less kinetic damping, this model can finally provide a closer match with the measurement than the Drude model. 

The pMRP allows simultaneous measurement of electron density, electron temperature, and electron-neutral collision frequency. Via the Drude model, we can calculate electron density from the measurement. This collision-less kinetic model offers an opportunity to calculate electron density and electron temperature in the low-pressure plasma. In the future, we will include collisions in the kinetic model, which can be eventually used to obtain electron density, electron temperature, and electron-neutral collision frequency from the measurement. This model will be valid under arbitrary pressure.

\pagebreak

\section{Acknowledgments}

The authors gratefully acknowledge the financial support by Deutsche Forschungsgemeinschaft (DFG) via the project DFG 360750908. Gratitude is expressed to the MRP-Team at Ruhr University Bochum.

\renewcommand\refname{\fontsize{11pt}{13pt} \selectfont References \\ \vspace*{-22pt}}
\begin{spacing}{1} 

\end{spacing}

\pagebreak

\section*{Figures}
\iffigures
\begin{figure}[h!]
	\centering
	\includegraphics[width=0.8\textwidth]{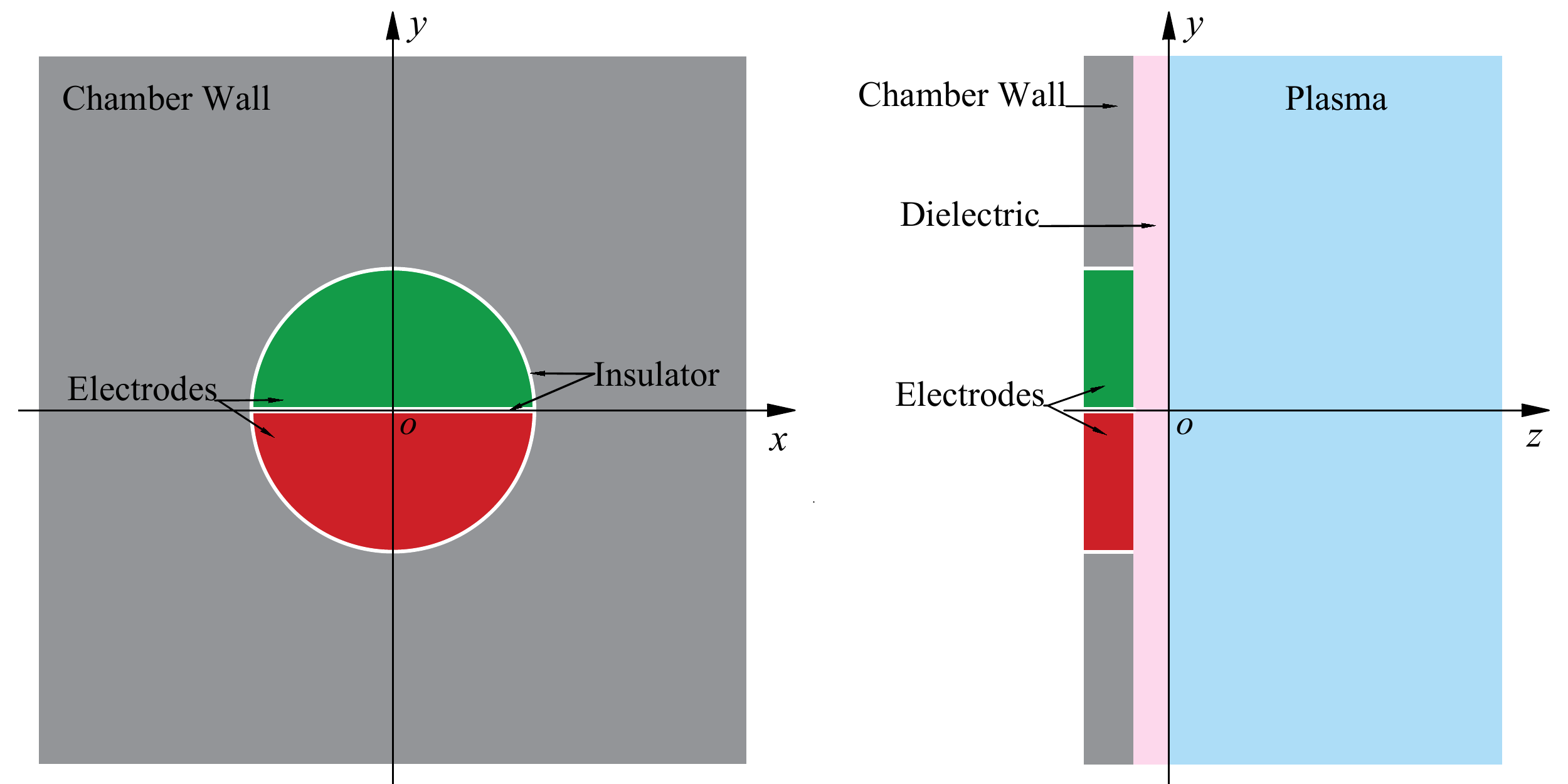} \\
	\caption{Idealized planar Multipole Resonance Probe. Two semi-disc electrodes $E_\pm$ with a radius of $R$ are flatly integrated into the chamber wall. The electrodes are insulated from each other and from the grounded chamber wall. A thin dielectric with a thickness of $d$ covers the electrodes and the chamber wall.}
	\label{Ideal pMRP}
\end{figure}

\begin{figure}[h!]
	\centering
	\includegraphics[width=0.8\textwidth]{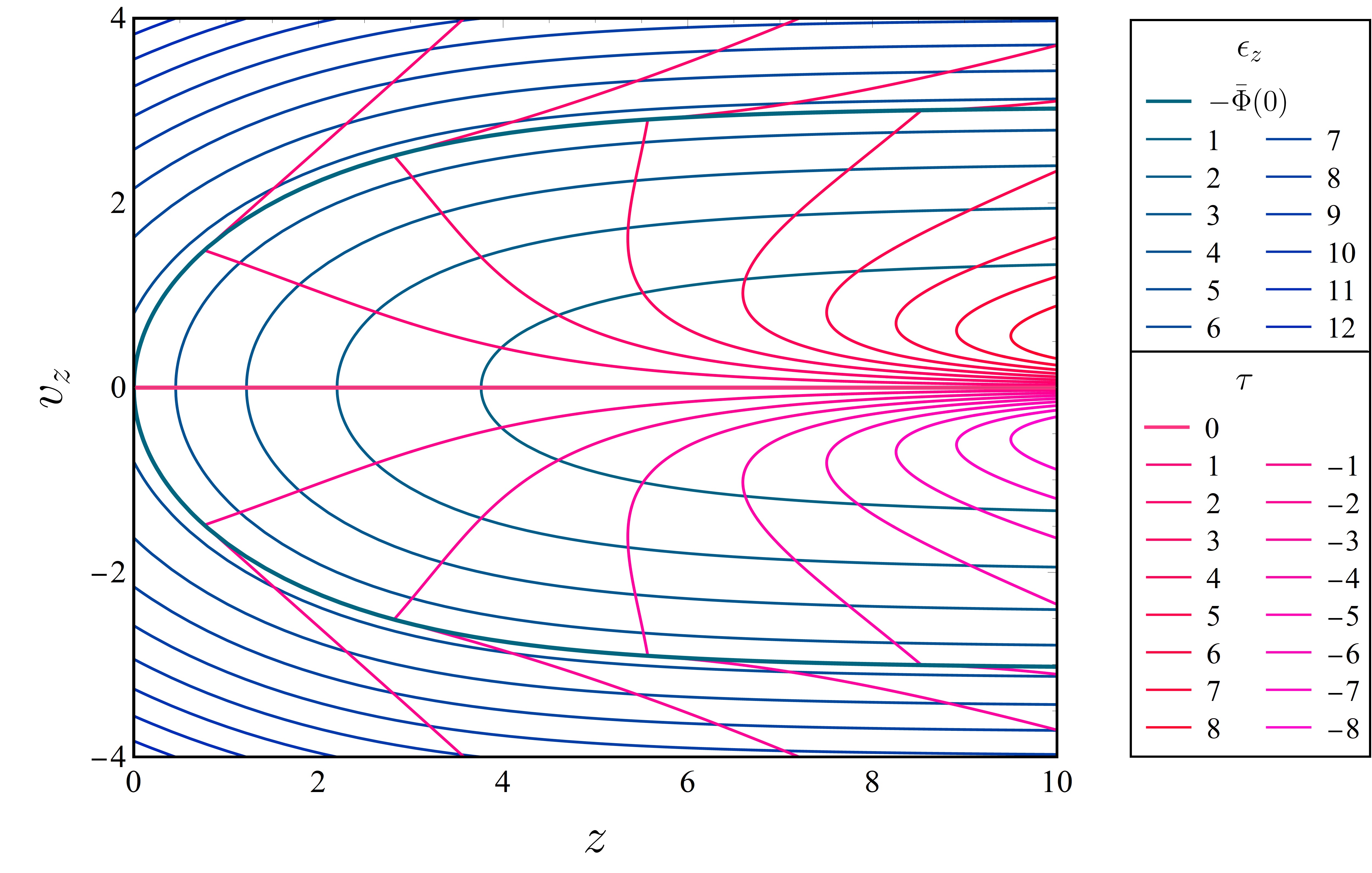} \\
	\caption{Unperturbed trajectory. The blue curve shows the variation of $(z,v_z)$ at a certain $\varepsilon_z$. The electron enters from infinity at an initial speed of $v_z=-\sqrt{2\varepsilon_z}$, then turns around at $z={\bar\Phi}^{-1}\left(-\varepsilon_z\right)$ ($\varepsilon_z<-\bar\Phi(0)$), or experiences a reflection at $z=0$ ($\varepsilon_z>-\bar\Phi(0)$), and finally leaves at a speed of $v_z=\sqrt{2\varepsilon_z}$. The bold blue curve represents the critical trajectory $(\varepsilon_z=-\bar\Phi(0))$. The red curve depicts the temporal parameter $\tau$ which always takes a turning when crossing the critical trajectory.}
	\label{Unperturbed trajectory}
\end{figure}
\pagebreak

\begin{figure}[h!]
	\centering
	\includegraphics[width=0.6 \textwidth]{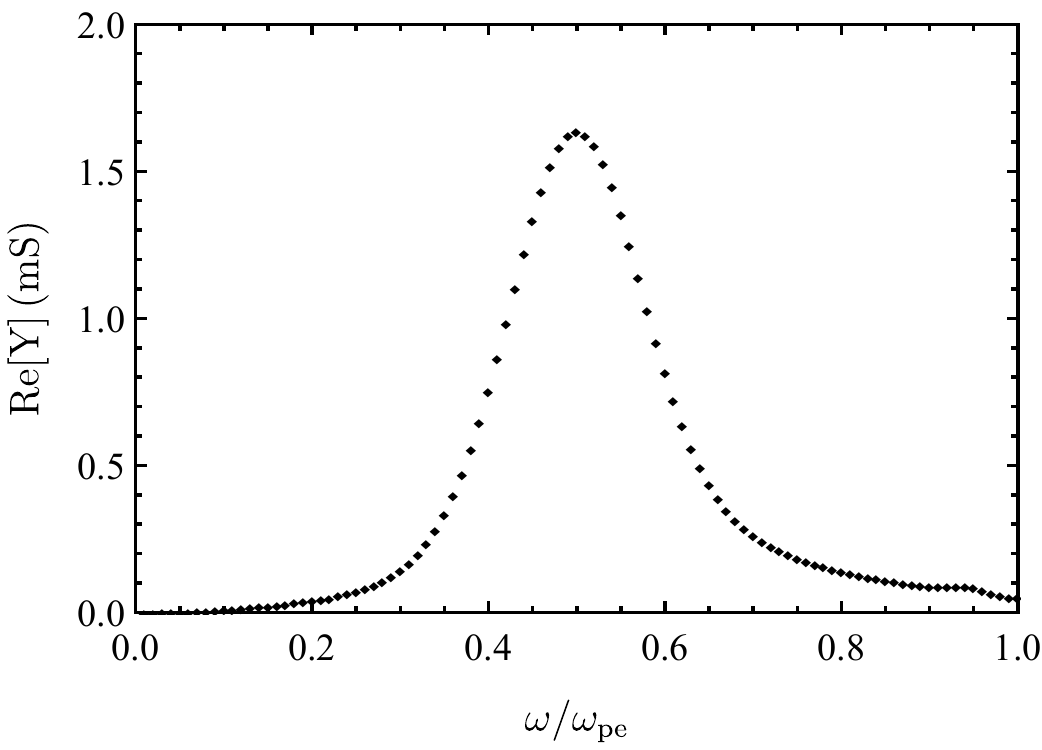} \\
	\caption{Spectrum of the collision-less kinetic model: $n_\mathrm{e}=4\times10^{15} \,\mathrm{m}^{-3},\,T_\mathrm{e}=2\,\mathrm{eV}$}
	\label{fig-kinetic}
\end{figure}

\begin{figure}[h!]
	\centering
	\includegraphics[width=0.6 \textwidth]{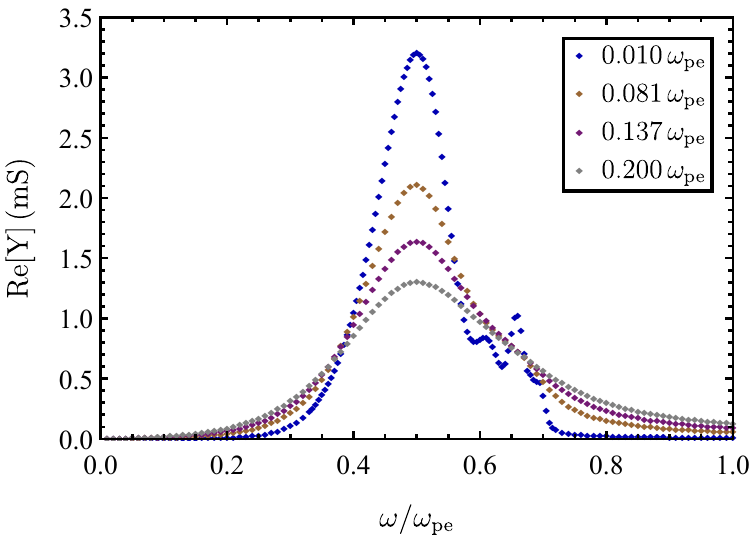} \\
	\caption{Spectra of the Drude model: $n_\mathrm{e}=4\times10^{15} \,\mathrm{m}^{-3},\,\delta=0.50\,\mathrm{mm},\,\nu=0.010\sim0.200\,\omega_\mathrm{pe}$}
	\label{fig-Drude}
\end{figure}
\pagebreak

\begin{figure}[h!]
	\centering
	\includegraphics[width=0.6 \textwidth]{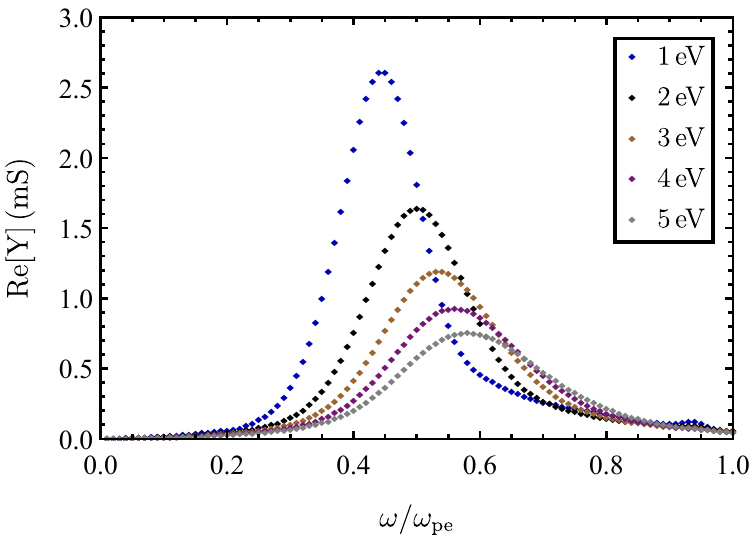} \\
	\caption{Spectra of the collision-less kinetic model: $n_\mathrm{e}=4\times10^{15} \,\mathrm{m}^{-3},\,T_\mathrm{e}=1\sim5\,\mathrm{eV}$}
	\label{fig-kinetic-Te}
\end{figure}

\begin{figure}[h!]
	\centering
	\includegraphics[width=0.6 \textwidth]{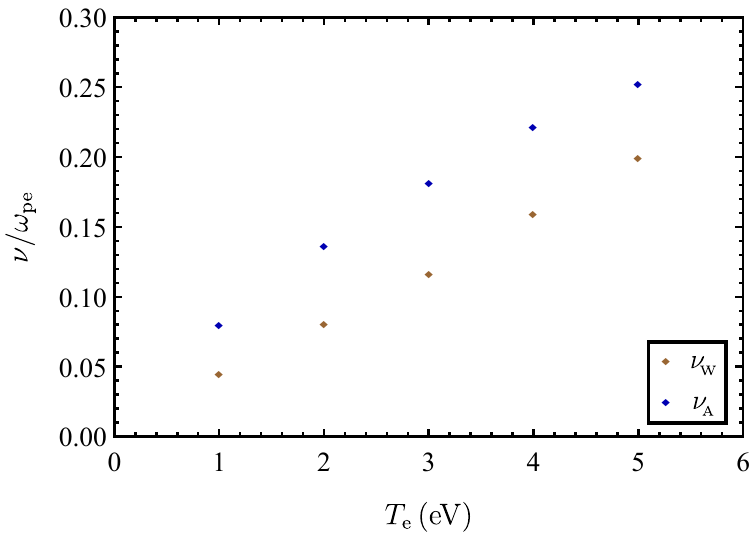} \\
	\caption{Collision frequencies $\nu_\mathrm{\!_W}$ and $\nu_\mathrm{\!_A}$: $n_\mathrm{e}=4\times10^{15} \,\mathrm{m}^{-3},\,T_\mathrm{e}=1\sim5\,\mathrm{eV}$}
	\label{fig-nu}
\end{figure}

\end{document}